\documentclass[nohyper, letterpaper,11pt,notoc]{JHEP3}

\usepackage{graphicx}
\usepackage{amssymb}
\usepackage{amsmath}
\usepackage{latexsym}
\usepackage{slashed}
\newcommand{\GeV}{~\mathrm{GeV}}

\title{On the Existence of Angular Correlations in Decays with Heavy Matter Partners}
\author{Can Kilic$^{(a)}$, Lian-Tao Wang$^{(b)}$ and Itay Yavin$^{(b)}$
\\ \it{(a) The Johns Hopkins University, Dept. of Physics and Astronomy, Baltimore MD 21218}
\\ \it{(b) Joseph Henry Laboratories, Princeton University, Princeton NJ 08544}}
\abstract{If heavy partners of the Standard Model matter fields are
discovered at the LHC it will be imperative to determine their spin
in order to uncover the underlying theory. In decay chains, both the
spin and the mass hierarchy of all particles involved can influence
the resulting angular correlations. We present the necessary
conditions for decays involving the matter partners to exhibit
angular correlations. In particular we find that when the masses are
not degenerate, a heavy fermionic sector always displays angular
correlations in cascade decays. When the masses are closely
degenerate the size of spin effects is controlled by a global phase
parameter. In a large region of the parameter space, correlations
are strongly suppressed. In other regions, such as in UED model where this phase is fixed by 5-d Lorentz invariance, the
correlations are pronounced. In addition, we show that in certain
cases one may even have enough information to determine the spin of
other heavy partners involved in the decay (such as the Lightest
Stable Partner or a heavy gluon).}

\begin{document}
\section{Introduction}

Stabilizing the hierarchy between the electroweak (EW) scale and the
Planck scale generically requires the existence of new particles at
the TeV scale. Typically, such new particles organize themselves
into partners of the Standard Model (SM) particles, i.e. they have
the same gauge quantum numbers. However, the spin of such partners
is more model dependent. There exist models where the partners' spin
is opposite to that of the SM particles (the prime example is of
course Supersymmetry). There are also several other models where the
partners' spin is the same as that of the SM particles (e.g. UED
\cite{Appelquist:2000nn,Cheng:2002ab} or Little Higgs
models\cite{Arkani-Hamed:2002qy,Arkani-Hamed:2002qx,Cheng:2003ju,Cheng:2004yc,Low:2004xc,Cheng:2005as,Cheng:2006ht})

Recently, numerous papers began to address the issue of direct spin
determination at the LHC. The comparison was usually made between
Supersymmetry (SUSY) and the Universal Extra-dimensions (UED)
scenario
\cite{Barr:2004ze,Battaglia:2005zf,Smillie:2005ar,Datta:2005zs,Datta:2005vx,Barr:2005dz,Alves:2006df}.
In this paper we would like to point out general features of any
such comparison which go beyond the particular model of UED (see
also \cite{Athanasiou:2006ef,Wang:2006hk,Smillie:2006cd}).

We begin by analyzing the case of fermionic partners of SM matter
fields (i.e. quarks and leptons) with a general mass spectrum and
model independent couplings. As we emphasized in a previous
publication \cite{Wang:2006hk} the chirality of the couplings is a
crucial ingredient in any attempt at spin determination. We give a
complete description of the necessary conditions for chiral
couplings of the new matter sector. These conditions depend strongly
on the spectrum. We identify the relevant parameters to be the
relative size of the splitting between the Dirac masses $\Delta M$
and the Yukawa mass $\sim \lambda v$. In the case where there is a
large splitting between the partners of the left and right handed SM
fields, the interactions are dominantly chiral.

However, when the matter partners' spectrum is closely degenerate,
the situation is more subtle. The chirality of the coupling is
controlled by a phase parameter in the fermionic mass matrix. It
interpolates between a limit where the chiral coupling is suppressed
and one where the chiral coupling is enhanced. For the UED model
this phase is fixed by 5-d Lorentz invariance and leads to an
enhanced chiral structure. However, in principle the situation could
be very different for a general model of fermionic partners of the
matter sector (e.g. Little Higgs). We examine the consequences of a
general phase structure on the prospects of spin determination.

In light of these observations it becomes clear that rather than
treating UED as a universal benchmark, studies of spin correlations
should extend to models with more general fermionic couplings. Due
to the close connection between chirality and 5-d Lorentz
invariance, this also means that experimental observation of
degenerate fermion partners with chiral couplings provides a unique
test of models based on extra dimensions.

\begin{figure}[!t]
\begin{center}
\includegraphics{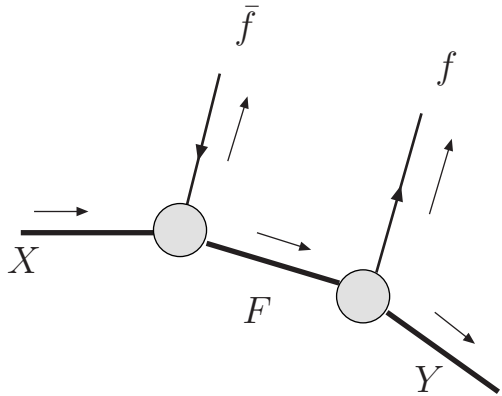}
\caption{The topology for a generic decay where spin information may
be found. A heavy particle $X$ decays into a SM fermion $\bar{f}$
and a heavy partner $F$ which subsequently decays into another SM
fermion $f$ and an invisible particle $Y$. The angular correlations
between $f$ and $\bar{f}$ may reveal the spin identity of the
intermediate particle $F$.} \label{fig:genericMab.eps}
\end{center}
\end{figure}

In addition, we investigate the possibility of extracting further
information from a decay involving the matter partners. We will
consider event topologies as shown in Fig.\ref{fig:genericMab.eps}
where $F$ is the matter partner. If $F$ is a scalar, Lorentz
invariance constrains $X$ and $Y$ to be fermions. Since there is no
correlation between the outgoing $f-\bar{f}$ to begin with, one
cannot hope to extract any more information. In contrast, when $F$
is a fermion, $X$ and $Y$ can be either a scalar or a vector-boson.
We discuss the dependence of the $f-\bar{f}$ angular correlation on
the spin of $X$ and $Y$ and show that, with sufficient information
regarding the mass spectrum, one can unambiguously determine the
spin of $X$ and $Y$.

The paper is organized as follows: In section \ref{sec:Chirality} we
lay down the necessary conditions for having chiral vertices in the
heavy matter sector with fermionic partners of the SM and in section
\ref{sec:HeavyGluon} we discuss the resulting angular correlations
in cascade decays involving the SM matter partners. In section
\ref{sec:OtherScenarios} we focus on additional information that can
be extracted out of such cascade decays. Section
\ref{sec:experimental} contains a brief discussion of some
experimental difficulties such measurements must confront. In
section \ref{sec:Conclusions} we present our conclusions.

\section{Chirality Conditions}
\label{sec:Chirality}

In this section we discuss the conditions needed for a heavy
fermionic sector to have chiral couplings with the SM fields. We
will assume that this new sector has the same quantum numbers as the
known quarks and leptons. In addition, we will assume that some
$Z_2$ parity symmetry is present (e.g. $KK$-parity in the UED model
or $T$-parity in Little Higgs models). We will use the quark sector
to demonstrate our results, but it should be clear that identical
conclusions apply to the lepton sector as well.

\begin{table}[h]
\begin{center}
\begin{tabular}{c|c|c}
SM & Heavy partners  &$SU(2)\times U_Y(1)$  \\
\hline $q_L$ & \parbox{5cm}{$Q_L^\prime = \left(
\begin{array}{l}
u_L^\prime \\
d_L^\prime \\
\end{array}
\right) \quad Q_R^\prime = \left(
\begin{array}{l}
u_R^\prime \\
d_R^\prime \\
\end{array}
\right)$} & $\left(2,\frac{1}{6}\right)$
\\
$u_R$ & $U_R^\prime \hspace{2.2cm} U_L^\prime$ &
$\left(1,\frac{2}{3}\right)$
\\
$d_R$ & $D_R^\prime \hspace{2.2cm} D_L^\prime$ &
$\left(1,-\frac{1}{3}\right)$
\end{tabular}
\end{center}
\end{table}
We denote the heavy fermionic modes by $Q_{L,R}^\prime$,
$U_{L,R}^\prime$ and $D_{L,R}^\prime$ (for simplicity we are
considering only one generation). Here, $Q^\prime =
(u^\prime,d^\prime)$ is an $SU(2)$ doublet while $U^\prime$ and
$D^\prime$ are singlets; note that $Q_L^\prime$,$U_R^\prime$ and
$D_R^\prime$ are partners to SM fermions while
$Q_R^\prime$,$U_L^\prime$ and $D_L^\prime$ have no low energy
counterparts.

Due to the $Z_2$ parity, the heavy fermions can only couple to the
SM via heavy bosons ($g^\prime$, $Z^\prime$ and etc.). The chiral nature
of the SM restricts the form of the interactions. For example, the
couplings to the heavy gluon $g^\prime$ is
\begin{equation}
\label{eqn:interactions}
\mathcal{L}_{int} = \overline{Q}^\prime_L \slashed{g}^\prime q_L +
\overline{U}^\prime_R \slashed{g}^\prime u_R + \overline{D}^\prime_R \slashed{g}^\prime
d_R + h.c.
\end{equation}
where $q_L$ is the SM electroweak doublet and $u_R$ and $d_R$ are the singlets.

For the new fermions to be parametrically heavy they must have Dirac
masses of the form $M_Q \overline{Q}_L^\prime
Q_R^\prime$\footnote{In principle one can crank up the mass by
increasing the Yukawa coupling. However, such large Yukawa couplings
are limited by perturbativity and severely constrained by
measurements of the S parameter.}. Therefore, after EWSB their mass
matrix is given by,
\begin{equation}
\mathcal{L}_{mass,up} = \left(
\begin{array}{cc}
u^\prime_L & U^\prime_L
\end{array}
\right)
\left(
\begin{array}{cc}
M_Q & \lambda v \\
\lambda v & M_U~e^{i\varphi}
\end{array}
\right)
\left(
\begin{array}{c}
u^\prime_R \\
U^\prime_R
\end{array}
\right).
\label{eqn:massmatrix}
\end{equation}
where $v=246\GeV$, $\lambda$ is a Yukawa coupling which in general
is different than the corresponding coupling in the SM, and
$\varphi$ is some phase which cannot be rotated away. A similar mass
matrix holds for the down sector.

The diagonalization of this matrix is trivial, but it carries
important consequences to the prospects of spin measurements. The
mass eigenstates are a mixture of $u^\prime$ and $U^\prime$ given
by,
\begin{equation}
\left(
\begin{array}{c}
U_{1~(L,R)}^\prime \\
\\
U_{2~(L,R)}^\prime
\end{array}
\right)
=
\left(
\begin{array}{cc}
\cos\theta & e^{i\varphi^\prime}\sin\theta
\\
\\
-e^{-i\varphi^\prime}\sin\theta & \cos\theta
\end{array}
\right)
\left(
\begin{array}{c}
u_{(L,R)}^\prime\\
\\
U_{(L,R)}^\prime
\end{array}
\right)
\end{equation}
The phase $\varphi^\prime$ is given by,
\begin{equation}
\varphi^\prime = \tan^{-1}\left(\frac{M_U\sin\varphi}{M_Q + M_U\cos\varphi} \right)
\end{equation}
and the mixing angle is determined by the ratio,
\begin{equation}
\tan{\theta}=\frac{2\lambda v \sqrt{M_Q^2 + 2M_QM_U \cos\varphi + M_U^2}}{\left(M_Q^2-M_U^2\right) + \sqrt{\left(M_Q^2-M_U^2\right)^2 + 4\lambda^2v^2\left(M_Q^2 + 2M_QM_U\cos\varphi + M_U^2\right)}}
\label{eqn:rotationangles}
\end{equation}

We will be mostly interested in the two extreme cases where
$\varphi=0$ or $\varphi=\pi$ since the general case is a simple
interpolation in between. In these limiting cases the expression for
$\tan{\theta}$ simplifies and it is easy to see the conditions for
large or small mixing.

\subsection{Non-degenerate Spectrum}

For a non-degenerate spectrum $M_Q - M_U \gg \lambda v$, the mixing is always small and the phase plays no role,
\begin{equation}
\label{eqn:non-deg-mix}
\tan{\theta} \sim \frac{\lambda v}{\Delta M} \quad \ll \quad 1
\end{equation}
where $\Delta M = M_Q - M_U$. When the mixing is small the
interactions with the SM in Eq.(\ref{eqn:interactions}) are still
chiral after rotation into the mass eigenstate basis. This simple
observation leads to an important conclusion: any model with heavy
fermionic partners of the SM matter sector protected by some $Z_2$
parity ($KK$-parity, $T$-parity etc.), with a non-degenerate
spectrum, exhibits angular correlations in decays.

\subsection{Degenerate Spectrum}

We now turn to examine a degenerate spectrum, so that $M_Q - M_U \ll
\lambda v$. Such a spectrum can result from any symmetry that
relates left and right mass parameters. In this case the phase
$\varphi$ is important. When $\varphi=0$ the mixing is large,
\begin{equation}
\tan{\theta} = \frac{\lambda v}{\Delta M + \sqrt{\Delta M^2 + \lambda^2 v^2}} \rightarrow 1 - \frac{\Delta M}{\lambda v}
\end{equation}
The coupling of the mass eigenstates to the SM is no longer chiral,
\begin{equation}
\label{eqn:nonChiralInt}
\mathcal{L}_{int} = \overline{U}_1^\prime \slashed{g}^\prime u + \overline{U}_2^\prime \gamma_5 \slashed{g}^\prime u \quad +\quad  \mathcal{O}\left(\frac{\Delta M}{\lambda v}\right) + h.c.
\end{equation}
and similarly for the other gauge couplings. The mass splitting
between the two eigenstates is $\sim 2\lambda v$. If $\lambda v$ is
larger than the width of $U_{1,2}^\prime$ then in the narrow-width
approximation (NWA) there is no interference between diagrams
involving $U_1^\prime$ and those involving $U_2^\prime$. Therefore,
since the interactions in Eq.(\ref{eqn:nonChiralInt}) are not chiral
we expect to see no angular correlations in decays (up to
corrections of order $\mathcal{O}(\Delta M/\lambda v)$).

In the UED model, not only are the masses degenerate by
construction, but also the phase is fixed by 5-d Lorentz invariance
$\varphi = \pi$ as we show in appendix \ref{app:3-site}. In this
sense, UED is only a very special model of fermionic partners. In
this case, unless $\lambda$ is unnaturally large, the mixing is
always small,
\begin{equation}
\tan{\theta} = \frac{\lambda v}{\overline{M} +
\sqrt{\overline{M}^2+v^2\lambda^2}} \rightarrow \frac{\lambda
v}{2\overline{M}} \quad \ll \quad 1
\end{equation}
where $\overline{M} = (M_Q+M_U)/2$. Therefore, as can be seen from
Eq.(\ref{eqn:interactions}), the interactions of the SM with
$U_1^\prime$ and $U_2^\prime$ remain chiral. This will lead to
definite angular correlations in cascade decays.

The general case for an arbitrary phase is a simple interpolation
between these two extreme examples and we comment on it briefly
below.

\section{Angular Correlations in Heavy Vector-Boson Decay}
\label{sec:HeavyGluon}

In this section we illustrate the effect of the mixing between the
mass eigenstates on angular correlations in cascade decays. We
consider the decay of a heavy gauge partner into SM fermions and an
invisible particle. In SUSY, the decay of a gluino into a
$f$-$\bar{f}$ pair\footnote{In this section and the next we use $f$
to denote any of the SM fermions such as the quarks or leptons. We
denote by the letters $Q$ and $U$ quantities related to the $SU(2)$
doublet and singlet fields respectively.} through an on-shell
sfermion for example, contains no correlation between the outgoing
$f$-$\bar{f}$ pair because of the scalar nature of the intermediate
sfermion. However, in any scenario with partners of the same-spin as
the SM particles a correlation between the $f$-$\bar{f}$ pair may
exist. The existence of such correlations requires the interactions
to be chiral \cite{Wang:2006hk}. As we saw in the previous section
the chirality of the interactions depends on the mixing between the
heavy matter partners.

We begin by considering the amplitude for a gluon partner decay in same-spin scenarios,
given by
\begin{eqnarray*}
\label{eqn:Mab}
\mathcal{M} &=& \mathcal{M}_{1}+\mathcal{M}_{2} \\
&=&
\vcenter{\includegraphics[scale=0.7]{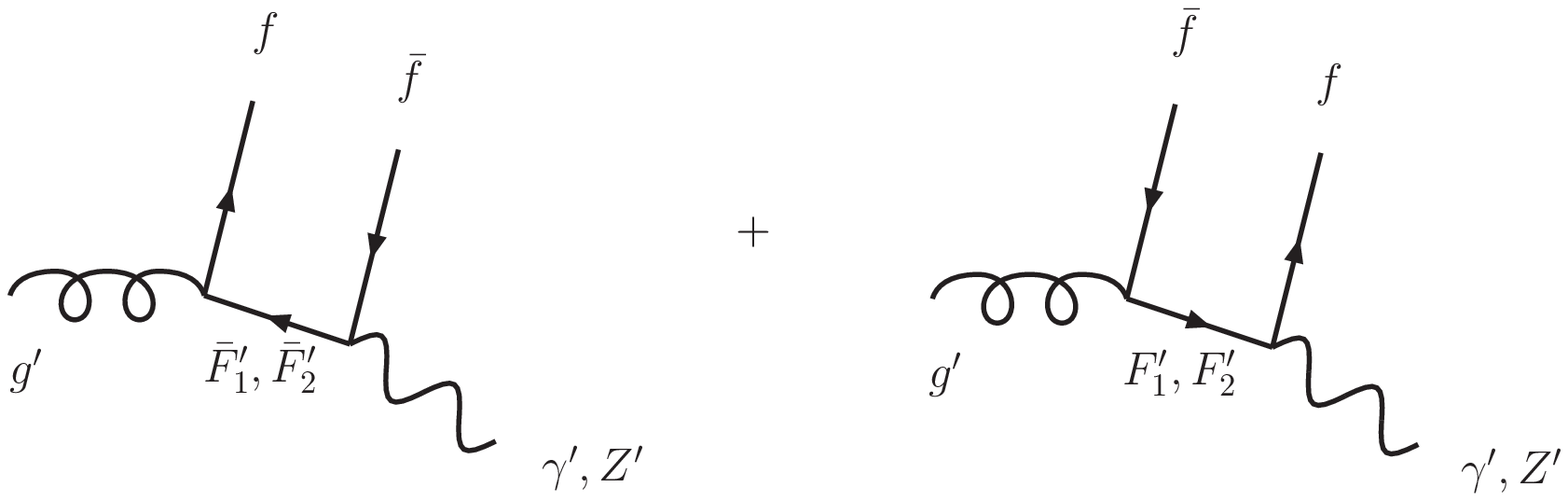}}
\end{eqnarray*}
The intermediate partner can be either of the mass eigenstates, $F_1^\prime$ or $F_2^\prime$.

If the intermediate particles are on-shell then the decay width
associated with this process can be written in the form
\begin{equation}
\label{eqn:parametrization} \frac{d\Gamma}{dt_{f\bar{f}}} \propto
\alpha + \beta ~ t_{f\bar{f}}
\end{equation}
where $t_{f\bar{f}} = (p_{f} + p_{\bar{f}})^2$ is the invariant mass
of the fermion pair\footnote{In the rest frame of the
fermion-partner, $t_{f\bar{f}} \propto (1-\cos\theta^*)$ where
$\theta^*$ is the angle between the $f-\bar{f}$ pair in that
frame.}. To emphasize the physically relevant effects we will only
display the spectrum-dependent part of $\alpha$ and $\beta$. As
usual, the invariant mass has a kinematical edge at,
\begin{equation}
t_{f\bar{f}}^{(edge)} = \frac{\left(M_{g^\prime}^2 - M_{int}^2 \right)  \left(M_{int}^2 - M_{\gamma^\prime}^2 \right)}{M_Q^2}
\end{equation}
where $M_{int}$ is the mass of the intermediate particle. Whether
the edge is actually visible or not depends on the slope.

In the case of SUSY partners, the slope always vanishes,
$\beta_{SUSY} = 0$, and no angular correlations are present.
Therefore to determine that the intermediate particle is not a
scalar, it is sufficient to determine that $\beta\ne 0$\footnote{In
general, for an intermediate particle of spin $s$, the differential
decay width is a polynomial in $t_{f\bar{f}}$ of degree $2s$.}. The
size of the slope $\beta$ depends on the chirality of the
interactions.

\subsection{Non-degenerate Spectrum $\Delta M \gg \lambda v$}

When the splitting in the diagonal elements of the mass matrix in
Eq.(\ref{eqn:massmatrix}) is large, the mixing between the two mass
eigenstates is minimal as illustrated in Eq.(\ref{eqn:non-deg-mix}).
In this case, the interactions with the SM are almost purely chiral.
Using the same parametrization as in Eq. (\ref{eqn:parametrization})
and working to leading order in $\frac{\lambda v}{\Delta M}$ we find
\begin{eqnarray}
\label{eqn:nonDegAmp}
\alpha_{F_1^\prime}&=& (M_{Q}^4 + 4 M_{\gamma^\prime}^2 M_{g^\prime}^2) ~t_{f\bar{f}}^{(edge)}
\nonumber\\
\nonumber\\
\beta_{F_1^\prime}&=& (2M_{g^\prime}^2-M_Q^2 )(M_Q^2-2M_{\gamma^\prime}^2)
\end{eqnarray}
Similar expressions hold for $\alpha_{F_2^\prime}$ and
$\beta_{F_2^\prime}$ with the replacement $M_Q \rightarrow M_U$.
Here, $F_1^\prime$ and $F_2^\prime$ represent the two mass
eigenstates forming the fermionic partners of the SM fermion $f$.

\vspace{7mm}
\begin{figure}[h]
\begin{center}
\vspace{1mm}
\includegraphics[scale=0.3]{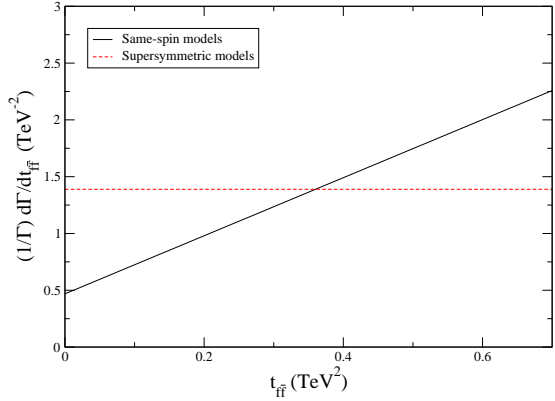}
\caption{A plot of $\frac{1}{\Gamma}\frac{d\Gamma}{dt_{f\bar{f}}}$ as a function of $t_{f\bar{f}}$ for the reaction shown in Eq.(\ref{eqn:Mab}). The same-spin scenario (solid-black) is contrasted with a supersymmetric scenario (dashed-red). The mass ratios are $M_{g^\prime}/M_F = 2$ and $M_F/M_{\gamma^\prime} = 5$. The only feature to note is simply the linear dependence of the same-spin scenario given by Eq.(\ref{eqn:nonDegAmp})}
\label{fig:theory_curves}
\end{center}
\end{figure}

There is no parametric suppression of the slope as the coefficient
of $t_{f\bar{f}}$ is of the same order as the constant term.
However, a kinematical suppression is still present and when
$M_{Q}^2/M_{\gamma^\prime}^2 =2$ the slope vanishes. The origin of
this suppression is simple: when $M_{Q}^2/M_{\gamma^\prime}^2 =2$,
the vector-boson $\gamma^\prime$ is an equal mixture of the transverse
and longitudinal components and spin conservation does not choose
any preferred alignment for the outgoing $f-\bar{f}$ pair. The sign
of the slope can therefore provide additional kinematical
information. As we shall see in the next section, it may, in fact,
help determine the spin of $g^\prime$ and $\gamma^\prime$ as well.

To conclude, in the absence of any kinematical suppression such as
$M_{Q}^2/M_{\gamma^\prime}^2 \sim 2$,  angular correlations are very
pronounced and a histogram of the events should readily discover the
linear dependence of $d\Gamma/dt_{f\bar{f}}$ on $t_{f\bar{f}}$.

\subsection{Nearly Degenerate Spectrum, $\Delta M \ll \lambda v$}

If the splitting between the diagonal elements of the mass matrix is
very small $\Delta M \ll \lambda v$ the phase difference $\varphi$
between the diagonal elements is important. While the expression for
the amplitude can be easily computed for a general phase, it is more
illuminating to consider the two extreme cases $\varphi=0$ and
$\varphi = \pi$.

When $\varphi=0$, mixing is maximal as we saw in the previous
section. The interaction of $F_1^\prime$ with the SM is almost
purely vector-like and that of $F_2^\prime$ is almost purely axial
as seen from Eq.(\ref{eqn:nonChiralInt}).

The mass splitting between the two states $F_1^\prime$ and
$F_2^\prime$ is approximately $2\lambda v$. If the splitting is
larger than the decay width of $F_1^\prime$ and $F_2^\prime$ then
the interference between the two diagrams in
Eq.(\ref{eqn:nonChiralInt}) can be neglected and the NWA is
justified\footnote{If the NWA is not valid, one must take the
interference into account. In this case the coefficients $\alpha$
and $\beta$ are given by similar expressions to
Eq.(\ref{eqn:nonDegAmp}).}. Using the NWA we find (to leading order
in $\Delta M/\lambda v$),
\begin{eqnarray}
\label{eqn:DegAmp}
\alpha &=& \frac{1}{2}
\left(2M_{g^\prime}^2+\overline{M}^2\right)\left(\overline{M}^2+2M_{\gamma^\prime}^2\right)~
t_{f\bar{f}}^{(edge)}
\nonumber\\
\nonumber\\
\beta
&=&\left(2M_{g^\prime}^2-\overline{M}^2\right)\left(\overline{M}^2-2
M_{\gamma^\prime}^2\right)~\left(\frac{\Delta M}{\lambda
v}\right)^{2}.
\end{eqnarray}
This expression is very similar to the one we found in the previous
case only that the coefficient of $t_{f\bar{f}}$ is subdominant to
the constant piece and comes only at second order in the small
parameter $\Delta M/\lambda v$. Angular correlations are therefore
suppressed and vanish altogether in the degenerate limit. The reason
for this is simple: both interaction vertices in Eq.(\ref{eqn:Mab})
must be at least partially chiral to have any angular correlations
\cite{Wang:2006hk}.

A very rough lower bound on the number of events needed to determine
that a non-zero slope is statistically significant is simply,
\begin{equation}
N \gtrsim \frac{1}{(\Delta M/\lambda v)^4}
\end{equation}
Even when the suppression is only moderate, $\Delta M/\lambda v \sim
1/10$,  the number of events required becomes exceedingly large.

On the other hand, when $\varphi=\pi$ as in UED, the mixing is
minimal and the amplitude is the same as in the non-degenerate case,
Eq.(\ref{eqn:nonDegAmp}). Therefore, UED predicts that angular
correlations are present between the $f-\bar{f}$ pair.

In an interesting recent paper, the authors of \cite{Alves:2006df}
claim that failure to observe angular correlations between the
$f-\bar{f}$ pair establishes the existence of the gluino (albeit
indirectly). The above discussion shows that this need not be the
case. If the spectrum is non-degenerate, the correlations may simply
be kinematically suppressed (when $M_Q^2 \sim 2
M_{\gamma^\prime,Z^\prime}^2$). This possibility may be ruled-out by
a proper measurement of the spectrum. If the spectrum is nearly
degenerate, the phase $\varphi$ plays an important role. In this
case, the mixing between the mass eigenstates, given by
Eq.(\ref{eqn:rotationangles}), is approximately,
\begin{equation}
\tan\theta = \frac{\lambda v\cos\left(\varphi/2\right)}{\Delta M + \sqrt{\Delta M^2 + (\lambda v)^2 \cos^2(\varphi/2)}}\quad \rightarrow \quad 1 + \frac{\Delta M}{\lambda v \cos(\varphi/2)}
\end{equation}
which is large in general. Only when $\varphi\rightarrow \pi$ does
this expansion break down and mixing is diminished. Therefore, as we
saw above, we expect the angular correlations to be suppressed by
$\Delta M/\lambda v \cos(\varphi/2) $. Failing to observe angular
correlation is therefore not necessarily an indication that the
underlying model is SUSY. However, it is sufficient to rule-out the
UED model which predicts $\varphi = \pi$.

\subsection{Dilepton Correlations and the $Z^\prime$ Vertex}

In the discussion above we examined the angular correlations between
the SM fermions $f-\bar{f}$. Undoubtedly, the measurement of such
correlations in the leptonic sector is considerably less challenging
than the quark sector. One possible decay involving leptons is
$Z^\prime \rightarrow \ell^+~\ell^-~\gamma^\prime$ through an
intermediate heavy lepton. One may then question our initial choice
for the coupling of the partners to the SM fields,
Eq.(\ref{eqn:interactions}). For example, a heavy $Z^\prime\sim
W_3^\prime$ is likely to mix very little with a heavy
$\gamma^\prime$ and therefore its interactions with the matter
sector are strongly chiral to begin with. In this case, no amount of
mixing between the mass eigenstates can change the chiral structure
of the interactions.

Nonetheless, the analysis above is still relevant since as we argued
before, the chirality of only one vertex is insufficient to
guarantee spin effects in cascade decays. One must show that the
other vertices are at least partially chiral to ascertain the
existence of angular correlations.

One may also argue that the mixing between the lepton partners is
always suppressed since the Yukawa couplings are small. In this
paper we make no attempt at any general statements of this kind and
only remark that in any model in which this is true, angular
correlations between the outgoing $\ell^-\ell^+$ pair are indeed
present.

\section{Determining the Spin of $g^\prime$ and $Z^\prime/\gamma^\prime$}
\label{sec:OtherScenarios}

In cases where angular correlations are present, one can obtain more
information than just the spin of the intermediate particle. It may
even be possible to determine the spin of all the particles in a
cascade decay.

In Table \ref{tbl:diffspin} we present the slope,
$\beta_{F_1^\prime}$ in Eq.(\ref{eqn:parametrization}), for
different spin choices for the external particles $g^\prime$ and
$\gamma^\prime$ (similar expressions hold for $\beta_{F_2^\prime}$).
When the two external particles are scalars the slope is
unambiguously negative. In contrast, when the gluon partner
$g^\prime$ is a vector-boson and $\gamma^\prime$ is a scalar the
slope is unambiguously positive. These are simple consequences of
spin conservation. However, when $\gamma^\prime$ is a vector-boson,
the sign of the slope depends on whether $\gamma^\prime$ is
longitudinally dominated ($M_Q^2 > 2M_{\gamma^\prime}^2$) or
transversally dominated ($M_Q^2 < 2M_{\gamma^\prime}^2$).

Knowledge of the slope together with a measurement of the ratio
$M_Q/M_{\gamma^\prime}$ (possibly from kinematical edges) can
determine the spin of the external particle up to a two fold
ambiguity. For example, if we measure a positive slope and
$M_Q^2/M_{\gamma^\prime}^2 > 2$ we can conclude that the gluon partner
is a vector-boson, but we do not know whether $\gamma^\prime$ is a
scalar or a vector-boson. On the other hand, if the slope is
positive and $M_Q^2/M_{\gamma^\prime}^2 > 2$ we would have concluded
that $\gamma^\prime$ is a scalar, but we would have an ambiguity
left regarding the spin of the gluon partner.

\begin{table}[h]
\begin{center}
\begin{tabular}{c|c|c}
Scenario & Slope  $\beta$ & Intercept $\alpha$ \\
\hline
\includegraphics{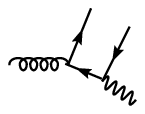} & $\left(2 M_{g^\prime}^2- M_Q^2 \right)\left(M_Q^2 - 2M_{\gamma^\prime}^2 \right)$ & $(M_{Q}^4 + 4 M_{\gamma^\prime}^2 M_{g^\prime}^2) ~t_{f\bar{f}}^{(edge)}$\\
\includegraphics{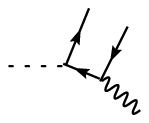} & $-\left(M_Q^2 - 2M_{\gamma^\prime}^2 \right)$ & $ M_Q^2 ~ t_{f\bar{f}}^{(edge)} $\\
\includegraphics{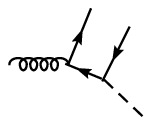} & $\left(2M_{g^\prime}^2 - M_Q^2 \right)$ & $M_Q^2 ~ t_{f\bar{f}}^{(edge)} $\\
\includegraphics{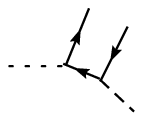} & $-1$ & $ t_{f\bar{f}}^{(edge)} $\\
\end{tabular}
\end{center}
\caption{If angular correlations exist between the outgoing
$f-\bar{f}$ or dilepton pair, then the sign of the slope of the
distribution (whether $\beta>0$ or $\beta <0$) may reveal the spin of the external particles
as well as the intermediate one. In the first row we consider a
scenario where the external particles are both vector-bosons. In the
second row the incoming partner is a scalar whereas the outgoing
partner is a vector-boson and so forth.} \label{tbl:diffspin}
\end{table}

\begin{figure}[h]
\begin{center}
\includegraphics{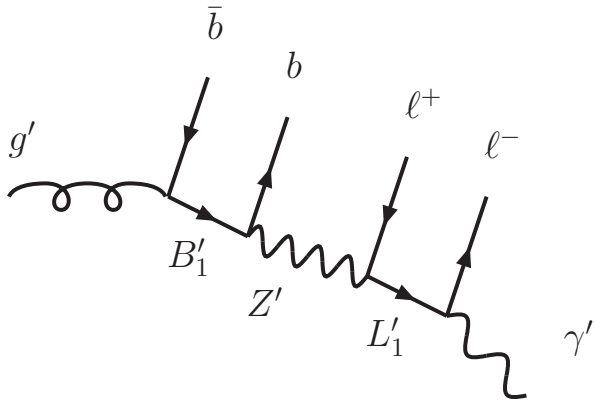}
\end{center}
\caption{A possible long cascade decay in scenarios with same-spin
partners. As explained in the text, a knowledge of the angular
correlations between the $b-\bar{b}$ pair and dilepton
$\ell^--\ell^+$ may contain enough information to determine the spin
of all the particles in the event, including $g^\prime$, $Z^\prime$
and $\gamma^\prime$.} \label{fig:long_cascade}
\end{figure}

Long cascade decays such as the one presented in Fig.
\ref{fig:long_cascade} may contain enough information to determine
the spin of all the partners unambiguously. For example, suppose we
measure the slope of the $f-\bar{f}$ pair to be negative with
$M_{F^\prime}^2/M_{Z^\prime}^2 < 2$ and that of the dilepton pair,
$\ell^--\ell^+$, to be negative with
$M_{L^\prime}^2/M_{\gamma^\prime}^2<2$ as well. Then, either all
three partners, $g^\prime$, $Z^\prime$ and $\gamma^\prime$, are
vector-bosons, or all three are scalars. Hence, with a single spin
measurement of the $Z^\prime$, such as described in
\cite{Barr:2004ze,Battaglia:2005zf,Smillie:2005ar,Datta:2005zs,Wang:2006hk},
we can lift this two-fold ambiguity and determine the spin of all
the particles in the event.

There are of course other discrepancies between the different
scenarios which can, in principle, help remove the degeneracies. For
example, in the limit where $M_{F_1^\prime} \gg M_{\gamma^\prime}$,
the diagram with a vector-boson $\gamma^\prime$ is longitudinally
enhanced over the other possibilities. However, these are numerical
differences that do not affect the overall shape and may be hard to
measure in practice when cross-sections and branching ratios cannot
be determined very accurately. We have tried to emphasize some
robust features of the distributions which do not relay on very
accurate determination of the shape.

\section{Experimental Challenges and Strategy}
\label{sec:experimental}

The most daunting experimental difficulty for spin determination
(and new physics in general) is SM background. We do not have much
to say about it over what has already been discussed in the
literature. Events with several final state leptons, hard jets and
large missing energy are very rare in the SM and may prove to be
strong signals of new physics. In what follows we will assume that
some non-zero set of such events has been isolated and that the SM
background is under control.

Once such a set is established one may begin to search for events
where dilepton or $b-\bar{b}$ final states are present. At this
stage it will be important to try and construct the event topology
and identify a set of events with topologies such as discussed
above. One may then look for angular correlations between dilepton
or $b-\bar{b}$ pairs. However, there may be some ambiguity in the
pairing (e.g. more then two leptons in an event or two leptons
coming from different branches) which will result in a reduction of
the signal due to combinatorics. As we showed in \cite{Wang:2006hk},
this is not necessarily a disaster and may simply require more
statistics to overcome the irreducible background.

\begin{figure}[h]
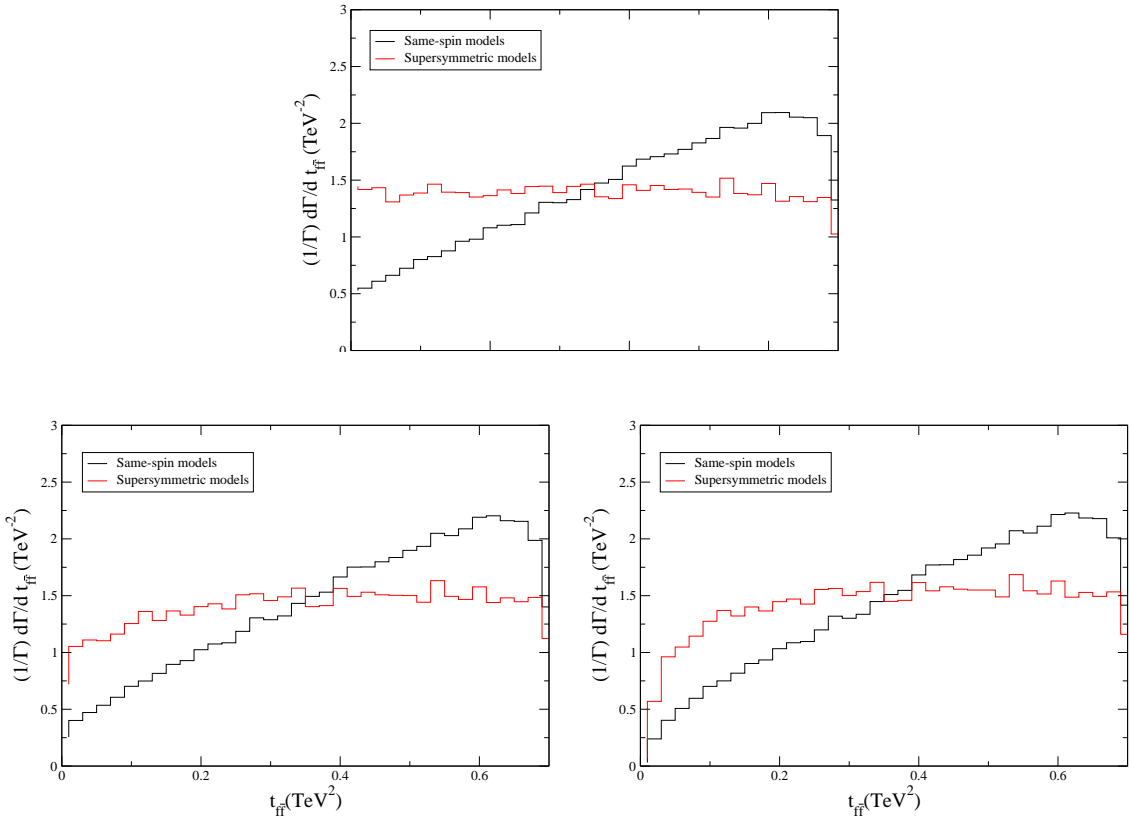

\begin{center}
\includegraphics[scale=0.3]{hist_nocuts.eps} \\\vspace{2mm}
\includegraphics[scale=0.3]{hist_DR03.eps}\hspace{3mm}
\includegraphics[scale=0.3]{hist_DR07.eps}
\caption{Plots of the decay distribution vs. $t_{f\bar{f}} =
(p_f+p_{\bar{f}})^2$ for several experimental cuts. In the upper
pane we present the Monte-Carlo results with no cuts. In the lower
left pane a $p_T>100\GeV$ cut was applied to both outgoing fermions
together with $\Delta R >0.3$ isolation cut. In the lower right
pane the isolation cut was increased to $\Delta R > 0.7$. The mass
parameters are such that $M_{g^\prime}/M_F =2$ and
$M_F/M_{\gamma^\prime} = 5$. The deterioration of the signal due to
the cuts is visible at the lower end of the distribution.}
\label{eqn:expcuts}
\end{center}
\end{figure}

An additional experimental aspect which may affect the signal is the
collection of various cuts imposed on the outgoing particles. In
particular $p_T$ and $\Delta R$ cuts will affect any invariant mass
distribution such as the one described above. To investigate this
point further we used the event generator HERWIG
\cite{Corcella:2002jc,Moretti:2002eu} which we modified to include
theories with same-spin partners \cite{Wang:2006hk}. In Fig.
\ref{eqn:expcuts} we plot a histogram of the decay distribution as a
function of $t_{f\bar{f}}$ for several experimental cuts. The
comparison is made between SUSY and a general same-spin scenario. We
require $p_T>100\GeV$ on both outgoing fermions. We also apply
different isolation cuts, $\Delta R> 0.3$ and $\Delta R >0.7$,
between the two fermions. While these cuts do not dramatically
affect the distribution they do modify its lower end.

These results are presented here with the intent to illustrate the
possible reshaping of such distributions due to experimental cuts.
Clearly, many more experimental challenges have to be overcome
before such a measurement is dubbed realistic. We emphasize that it
is important to base such measurements on robust features of the
distributions and not on tiny shape differences sensitive to a
variety of experimental factors which may not even be sufficiently
well-understood. Two such robust and important features to focus on
are the existence of a non-zero slope and its sign. \vspace{7mm}

\section{Conclusions}
\label{sec:Conclusions}

If new particles are discovered at the LHC, it will be important to
determine their spin. In light of the proliferation of new models in
the recent decade, it is evident that Supersymmetry is not the only
viable model for TeV scale physics. Many of the new models have, in
fact, partners to the SM particles with the same spin as their SM
counterparts. In this paper we investigated the general structure of
such a sector and described the necessary conditions for spin
effects to be observable.

When the fermionic partners of the left and right handed matter
fields ($q_L$ and $u_R$ and $d_R$) have well-separated masses,
angular correlations are always present regardless of the precise
model (UED, Little Higgs and etc.). However, as the spectrum is squeezed
and becomes more degenerate the mixing between the states is
sensitive to an arbitrary phase in the mass matrix. In the UED model
this phase is set by 5-d Lorentz invariance and the mixing remains
small. In more general models this need not be the case and mixing
can be very large. If such is the case, then spin information is
washed out and consequently becomes harder to observe.

An observation of a non-zero slope will reveal the matter partners
to be fermions. The sign of this slope may then be used to determine
the spin of the external particles in the decay (such as the heavy
gluon partner or other heavy vector-bosons). In certain cases it may
even by possible to unambiguously infer the spin of all the
particles involved in a cascade.

\vspace{0.25 in}
\textbf{Acknowledgments} : The Feynman diagrams in
this paper were generated using Jaxodraw \cite{Binosi:2003yf} and
the plots were produced with the help of R  \cite{Rprogram} and
Grace. The work of L.W. and I.Y. is supported by the National
Science Foundation under Grant No. 0243680 and the Department of
Energy under grant \# DE-FG02-90ER40542. The work of C.K. is
supported by the National Science Foundation under Grant
NSF-PHY-0401513 and the Johns Hopkins Theoretical Interdisciplinary Physics and Astrophysics Center.
Any opinions, findings, and conclusions or recommendations expressed in this material are those of the
author(s) and do not necessarily reflect the views of the National
Science Foundation.

\appendix
\renewcommand{\theequation}{A-\arabic{equation}}
\setcounter{equation}{2}
\section{3-site Deconstruction of UED}
\label{app:3-site}

In this appendix we present a simple 3-site toy model for a scenario
with same-spin SM partners. We illustrate the origin of the sign
difference between the KK mass of the doublet $Q^\prime$ and that of
the singlet $U^\prime$ (or $D^\prime$) in the UED model. We show
that it is simply a consequence of the assumed 5-d Lorentz
invariance of UED models. This serves to show that unless some
special symmetry is enforced there is, in general, a phase
difference between the masses, which is not necessarily $\varphi =
\pi$ as in UED. Therefore, as we argued in the text, UED can be
treated as a special limit of a generic 3-site model with same-spin
partners.

We begin by writing a simple lagrangian involving 3 pairs of $SU(2)$
doublets, $\{Q_{iL}(x),Q_{iR}(x)\}$ with $i=1,2,3$,
\begin{align}
\mathcal{L}_{Q} = \text{kinetic terms} +\frac{1}{a} \left(
\overline{Q}_{1R}, \overline{Q}_{2R}, \overline{Q}_{3R}\right)
\left(
\begin{array}{ccc}
0 & 1 & -1 \\
-1 & 0 & 1 \\
1 & -1 & 0
\end{array}
\right) \left(
\begin{array}{c}
Q_{1L}  \\
Q_{2L}  \\
Q_{3L}
\end{array}
\right)  + h.c.
\end{align}
This is simply a discrete version (with 3-sites) of an extra
dimension compacted on a circle with inter-site separation $a$. The
kinetic terms include the coupling to an $SU(2)$ gauge group on
every site \footnote{We are ignoring the gauge sector for
simplicity, although one must include hopping terms in that sector
as well.}.

As it stands, this lagrangian has a $Z_3$ symmetry which reshuffles
all three fields. This symmetry corresponds to the global $S_1$
symmetry, which is the conservation of 5-d momentum in the continuum
limit. This $Z_3$ is broken down to $Z_2$ by orbifolding the
geometry and identifying the first and third sites. This is most
easily seen by going to an eigenstate basis of $Z_2$ and rewriting
the lagrangian as,
\begin{align}
\label{eqn:CosSinBasis} \mathcal{L}_{Q} = \text{kinetic terms}
+\frac{1}{a} \left( \overline{Q}_{R}^{(0)},
\overline{Q}_{R}^{(cos)}, \overline{Q}_{R}^{(sin)}\right) \left(
\begin{array}{ccc}
0 & 0 & 0 \\
0 & 0 & -\sqrt{3} \\
0 & \sqrt{3} & 0
\end{array}
\right) \left(
\begin{array}{c}
Q_{L}^{(0)}  \\
Q_{L}^{(cos)}  \\
Q_{L}^{(sin)}
\end{array}
\right)  + h.c.
\end{align}
where we used the decomposition of the identity to write,
\begin{equation}
 \left(
\begin{array}{c}
Q_{1L}  \\
Q_{2L}  \\
Q_{3L}
\end{array}
\right) = \frac{1}{\sqrt{3}}\left(
\begin{array}{c}
1  \\
1  \\
1
\end{array}
\right) Q_L^{(0)} + \frac{1}{\sqrt{6}} \left(
\begin{array}{c}
1  \\
-2  \\
1
\end{array}
\right) Q_L^{(cos)} + \frac{1}{\sqrt{2}}\left(
\begin{array}{c}
1  \\
0  \\
-1
\end{array}
\right) Q_L^{(sin)}
\end{equation}
and similarly for the right-handed fermions. $Q^{(0)}$ is the zero
mode and $Q^{(cos)}$ and $Q^{(sin)}$ correspond to the first KK
level. These modes are even ($Q^{(0)}$,  $Q^{(cos)}$) and odd (
$Q^{(sin)}$) under the $Z_2$ symmetry. Orbifolding the geometry we
can project out the even (odd) states by choosing Dirichlet
(Neumann) boundary conditions,
\begin{align}
Q_1 = Q_3 \quad &\rightarrow \quad Q^{(sin)} = 0 \quad \quad \text{Neumann b.c.} \\
Q_2 =0 \quad Q_3 = -Q_1 \quad &\rightarrow \quad Q^{(0)} = Q^{(cos)}
= 0 \quad \text{Dirichlet b.c.}
\end{align}

The SM contains a left handed doublet, but no right
handed one so we should project out the odd modes for $Q_L$ and the
even modes for $Q_R$. The resulting lagrangian is,
\begin{equation}
\label{eqn:Qmass} \mathcal{L}_{Q} = \text{kinetic terms} + m
\overline{Q}_R^{(sin)} Q_L^{(cos)} + h.c.
\end{equation}
where $m = \sqrt{3}/a$. Doing the same exercise with the singlet
fields $U$ or $D$, we would have to choose the opposite boundary
conditions. We see from Eq.(\ref{eqn:CosSinBasis}) that we would
pick the opposite sign for the singlets,
\begin{equation}
\label{eqn:Umass} \mathcal{L}_{U} = \text{kinetic terms} - m
\overline{U}_R^{(sin)} U_L^{(cos)} + h.c.
\end{equation}
and similarly for the $D$. However, this sign depends on an
arbitrary choice we made for the relative sign between the kinetic
terms and the mass terms. 5-d Lorentz invariance fixes this sign to
be the same for the doublets and singlets. This choice is the reason
for the relative sign between the mass terms in Eq.(\ref{eqn:Qmass})
and Eq.(\ref{eqn:Umass}). If no such symmetry is present, the phase
is arbitrary and we have,
 \begin{equation}
 \mathcal{L}_{U} = \text{kinetic terms} + e^{i\phi} ~ m \overline{U}_R^{(sin)} U_L^{(cos)} + h.c.
\end{equation}
Adding the contribution from the mixing with the Higgs mode we
arrive at the mass matrix quoted in the text,
\begin{equation}
\mathcal{L}_{kk-mass} = \left(
\begin{array}{cc}
\overline{Q}_R^{(sin)}, \overline{U}_R^{(cos)}
\end{array}
\right) \left(
\begin{array}{cc}
m & \lambda v \\
\lambda v & e^{i\phi} m
\end{array}
\right) \left(
\begin{array}{c}
Q_L^{(cos)} \\
U_L^{(sin)}
\end{array}
\right)
\end{equation}

\bibliography{ref-gluedecay}
\bibliographystyle{unsrt}
\end{document}